\documentclass[aps,twocolumn,preprintnumbers,amsmath,amssymb,floats,citeautoscript]{revtex4-1}
\usepackage{graphicx}% Include figure files
\usepackage{dcolumn}% Align table columns on decimal point
\usepackage{bm}% bold math
\usepackage{color}
\usepackage{times}

\begin{document}

\title{Crystallization and vitrification of electrons in a glass-forming charge liquid}
%\title{Charge crystallization and vitrification\\
%in a glass-forming charge liquid}

\author{S.~Sasaki$^{1,*}$}
\author{K.~Hashimoto$^{1,*,{\dag}}$}
\author{R.~Kobayashi$^1$}
\author{K.~Itoh$^1$}
\author{S.~Iguchi$^1$}
%\author{I.~Oshima$^2$}
\author{Y.~Nishio$^2$}
\author{Y.~Ikemoto$^3$}
\author{T.~Moriwaki$^3$}
\author{N.~Yoneyama$^4$}
\author{M.~Watanabe$^5$}
\author{A.~Ueda$^6$}
\author{H.~Mori$^6$}
\author{K.~Kobayashi$^7$}
\author{R.~Kumai$^7$}
\author{Y.~Murakami$^7$}
\author{J.~M$\ddot{\rm{u}}$ller$^8$}
\author{T.~Sasaki$^1$}

\affiliation{
$^1$\mbox{Institute for Materials Research, Tohoku University, Aoba-ku, Sendai 980-8577, Japan}\\
$^2$\mbox{Department of Physics, Faculty of Science, Toho University, Funabashi, Chiba 274-8510, Japan}\\
$^3$\mbox{SPring-8, Japan Synchrotron Radiation Research Institute, Sayo, Hyogo 679-5198, Japan}\\
$^4$\mbox{Graduate Faculty of Interdisciplinary Research, University of Yamanashi, Kohu, Yamanashi 400-8511, Japan}\\
$^5$\mbox{Institute of Multidisciplinary Research for Advanced Materials, Tohoku University, Sendai 980-8577, Japan}\\
$^6$\mbox{The Institute for Solid State Physics, The University of Tokyo, Kashiwa, Chiba 277-8581, Japan}\\
$^7$\mbox{CMRC and Photon Factory, Institute of Materials Structure Science,} \mbox{High Energy Accelerator Research Organization (KEK), Tsukuba, Ibaraki 305-0801, Japan}\\
$^8$\mbox{Institute of Physics, Goethe-University Frankfurt, Max-von-Laue-Str. 1, 60438 Frankfurt(M), Germany}\\
$^*$These authors contributed equally to this work.\\
$^{\dag}$\rm{To whom correspondence should be addressed. E-mail: hashimoto@imr.tohoku.ac.jp (K.H.)}
}

%\date{\today}

%\pacs{}

\maketitle

{\bf 
Charge ordering (CO) is a phenomenon in which electrons in solids crystallize into a periodic pattern of charge-rich and charge-poor sites owing to strong electron correlations. This usually results in long-range order. In geometrically frustrated systems, however, a glassy electronic state without long-range CO has been observed. We found that a charge-ordered organic material with an isosceles triangular lattice shows charge dynamics associated with crystallization and vitrification of electrons, which can be understood in the context of an energy landscape arising from the degeneracy of various CO patterns. The dynamics suggest that the same nucleation and growth processes that characterize conventional glass-forming liquids guide the crystallization of electrons. These similarities may provide insight into our understanding of the liquid-glass transition.
}

%%%%%%%%%%%%%%%%%%%%%%FIG 1%%%%%%%%%%%%%%%
\begin{figure*}[t]
\includegraphics[width=0.8\linewidth]{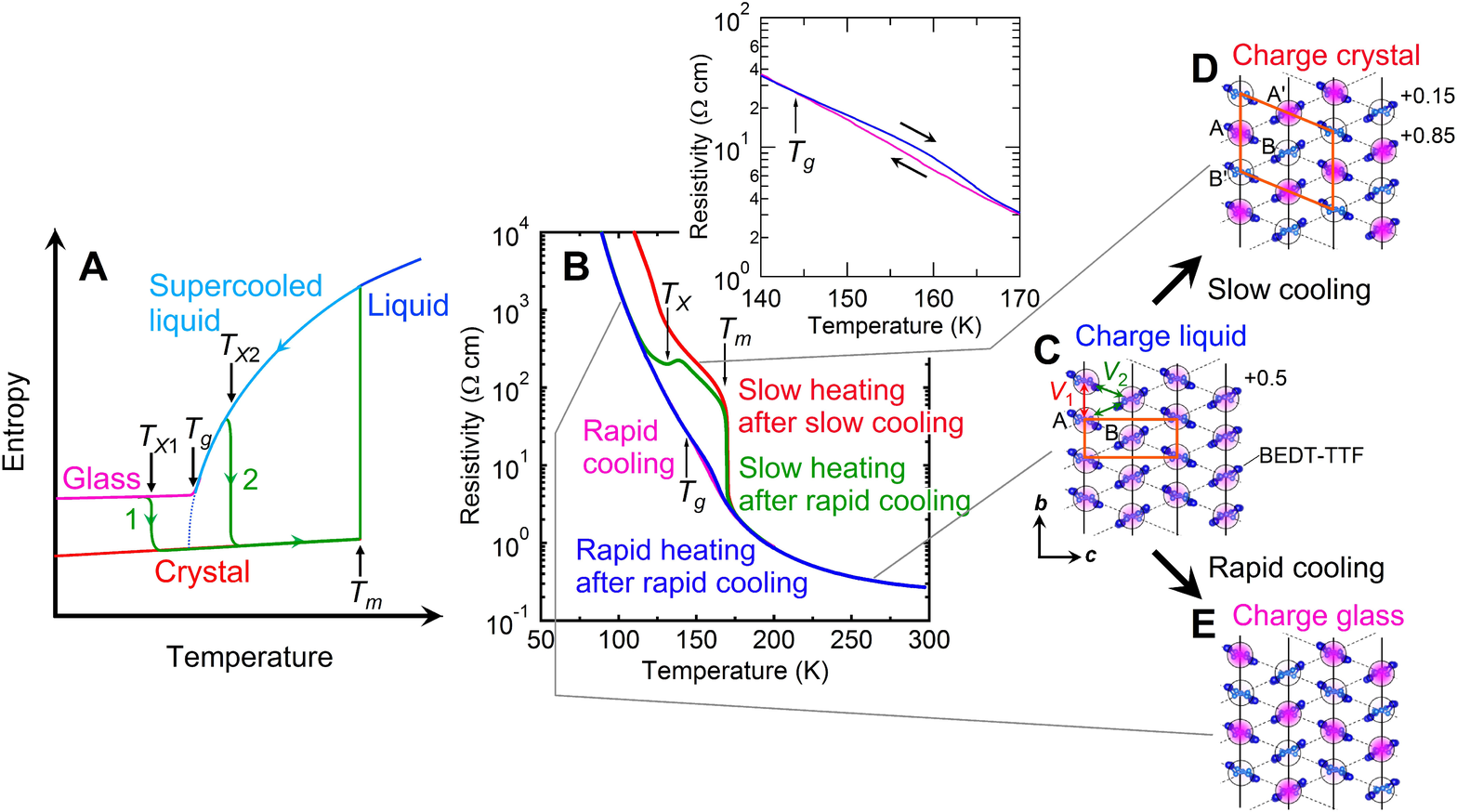}
\caption{ {\bf Charge crystallization and vitrification in $\theta_m$-(BEDT-TTF)$_2$TlZn(SCN)$_4$.} ({\bf A}) Generic entropy-temperature diagram of a glass-forming liquid. $T_m$ is the melting temperature, $T_g$ is the glass-transition temperature, and $T_X$ is the crystallization temperature. A supercooled liquid state emerges when the liquid is cooled quickly enough to avoid crystallization. Slower and faster heating processes (1 and 2) lead to a lower and higher crystallization temperatures ($T_{X1}$ and $T_{X2}$). ({\bf B}) Temperature dependence of the resistivity for $\theta_m$-TlZn measured in the rapid or slow heating process (after rapid or slow cooling) and in the rapid cooling process. Curves are color-coded. Charge crystallization occurs at $T_X$, which is lower than $T_g$, in the slow heating process after rapid cooling. Inset: Hysteresis loop of the resistivity during the heating/cooling process at a sweeping rate of 100 K/min. ({\bf C} to {\bf E}) Illustrations of the charge-liquid state (C), the charge-crystal state (D), and the charge-glass state (E) in $\theta_m$-TlZn. The orange lines in (C) and (D) denote the unit cell. At the CO transition, monoclinic cell is reduced to a triclinic cell ({\it{16}}). $V_1$ and $V_2$ are the nearest-neighbor Coulomb interactions, where $V_2/V_1 \sim 0.8$ (also see fig.\,S1D). The A and B sites in (C) are crystallographically equivalent, owing to the screw axis along the $b$ axis. In the charge-crystal state (D), a diagonal CO patter is formed, where the charge-rich (+0.85) sites, A and A', and the charge-poor (+0.15) sites, B and B', are not crystallographically equivalent. The difference in the charge density between A and A' (and between B and B') is too small to be detected experimentally ({\it{16}}).
} \label{optical}
\end{figure*}
%%%%%%%%%%%%%%%%%%%%%%FIG 1%%%%%%%%%%%%%%%

%1
The physics of glassy materials represents a fascinating problem in solid-state theory ({\it{1}}). Although progress has been made over the past several decades toward clarifying the dynamical aspects of the glass transition, the processes by which liquids acquire the glassy state upon cooling are not fully understood ({\it{2}}). The most fundamental nonequilibrium dynamic phenomena associated with the liquid-glass transition process are crystallization and vitrification. These phenomena are competing and mutually exclusive but are closely related to each other ({\it{3, 4}}). In glass-forming liquids, crystallization below the melting point $T_m$ can be avoided when the system is cooled quickly enough, leading to a supercooled liquid state accompanied by an increase in viscosity (Fig.\,1A) ({\it{2, 5}}). Upon further cooling, owing to the viscous retardation of crystallization, the supercooled liquid state freezes into a glassy state; that is, vitrification occurs at the glass transition temperature $T_g$. Thus, the relationship between crystallization and vitrification is key to the understanding of the liquid--glass transition. Here, we demonstrate that this general picture can be extended to crystallization and vitrification of strongly correlated electrons realized in a geometrically frustrated charge-ordered organic system, $\theta_m$-(BEDT-TTF)$_2$TlZn(SCN)$_4$ ({\it{6}}), where BEDT-TTF denotes bis(ethylenedithio)tetrathiafulvalene. Surprising similarities between our system and conventional glass formers are observed in the crystallization and vitrification processes, which highlight the universal nature of the liquid-glass transition.

%2. 
The quasi-two-dimensional (quasi-2D) organic materials $\theta$-(BEDT-TTF)$_2$$X$ consist of alternating stack of conducting BEDT-TTF and insulating anion $X$ layers; the BEDT-TTF molecules form a
triangular lattice ({\it{6--8}}). The charge transfer between these two layers leads to a 2D quarter-filled hole band system (that is, one hole per two BEDT-TTF molecules), in which the intersite Coulomb repulsions give rise to an instability towards charge ordering (CO) ({\it{9}}). Indeed, $\theta$-(BEDT-TTF)$_2$RbZn(SCN)$_4$ (henceforth $\theta$-RbZn), for example, undergoes a CO transition at 190 K ({\it{6, 7, 10, 11}}), where the charge carries are localized periodically with a horizontal stripe pattern (see the phase diagram in fig.\,S1C). Such a periodic CO state can be regarded as a ``charge-crystal" state ({\it{11}}). In contrast, above the CO transition temperature, the charge of +0.5 per one BEDT-TTF molecule is distributed uniformly in space; therefore, such a delocalized state can be referred to as a ``charge-liquid" state. 

In $\theta$-RbZn, when the sample is cooled faster than a critical cooling rate ($\sim$5 K/min), charge crystallization is kinetically avoided, leading to a ``charge-glass" state where the charge is randomly quenched. For comparison, in $\theta$-(BEDT-TTF)$_2$CsZn(SCN)$_4$ ($\theta$-CsZn), which has a more isotropic triangular lattice than $\theta$-RbZn, the critical cooling rate becomes much slower (fig.\,S1, C and D). As a result, the charge-liquid state inevitably results in a charge-glass state even upon very slow cooling ($<$ 0.1 K/min) ({\it{12}}). However, the mechanism of formation of the glassy electronic state---which has been discussed experimentally ({\it{11--14}}) and theoretically ({\it{15}}) in terms of the geometrically frustrated triangular lattice---still remains rather elusive. 

%3.\UTF{0081}\UTF{0081}\UTF{0081}\UTF{0084}\UTF{0081}%
The system $\theta$-(BEDT-TTF)$_2$TlZn(SCN)$_4$, which exists in two different crystal forms with orthorhombic and monoclinic symmetries (fig.\,S1, A and B) ({\it{6, 16}}), may play a key role in the understanding of the charge-glass state. Orthorhombic $\theta_o$-(BEDT-TTF)$_2$TlZn(SCN)$_4$ ($\theta_o$-TlZn), which has the same structural symmetry as $\theta$-RbZn and $\theta$-CsZn, exhibits a CO transition with a horizontal charge modulation at 240 K ({\it{16}}). Because the anisotropy of the triangular lattice is large, the critical cooling rate for glass formation remains quite high ({\it{13, 14}}). In contrast, the monoclinic $\theta_m$-(BEDT-TTF)$_2$TlZn(SCN)$_4$ ($\theta_m$-TlZn) shows a CO transition with a diagonal charge modulation at $T_m = 170$ K (Fig.\,1, B to D) ({\it{16}}). The different CO patterns of the two systems can be related to a difference in the strength of electron-lattice coupling, as pointed out by theoretical studies ({\it{17}}). Because the lattice distortion of $\theta_m$-TlZn at the CO transition is much smaller than that of $\theta_o$-TlZn, and because most of the entropy change of $\theta_m$-TlZn is of electronic origin (fig.\,S2), $\theta_m$-TlZn more likely approximates a system where the observed effects are purely electronic in nature. This is consistent with the extended Hubbard model (EHM) in the absence of electron-phonon coupling, in which the diagonal CO pattern rather than the horizontal one is expected ({\it{17}}).

The temperature-dependent resistivity $\rho(T)$ of $\theta_m$-TlZn shows a strong dependence on sweeping rate below $T_m$ (Fig.\,1B), although the triangular lattice for $\theta_m$-TlZn is more anisotropic than that for $\theta_o$-TlZn. In $\theta_m$-TlZn, the long-range CO transition can be avoided by rapid cooling ($\geq$ 50 K/min), and charge vitrification occurs through a supercooled charge-liquid state (Fig.\,1, B and E). In the cooling/heating cycle of the $\rho$-$T$ profile, a clear hysteresis loop associated with the glass transition is observed at 145 to 165 K (Fig.\,1B, inset), quite similar to what is observed in $\theta$-CsZn ({\it{12}}). We tentatively define the glass transition temperature $T_g$ as the temperature above which the resistivity starts to branch off.

%%%%%%%%%%%%%%%%%%%%%%FIG 2%%%%%%%%%%%%%%%
\begin{figure*}[tb]
\includegraphics[width=0.8\linewidth]{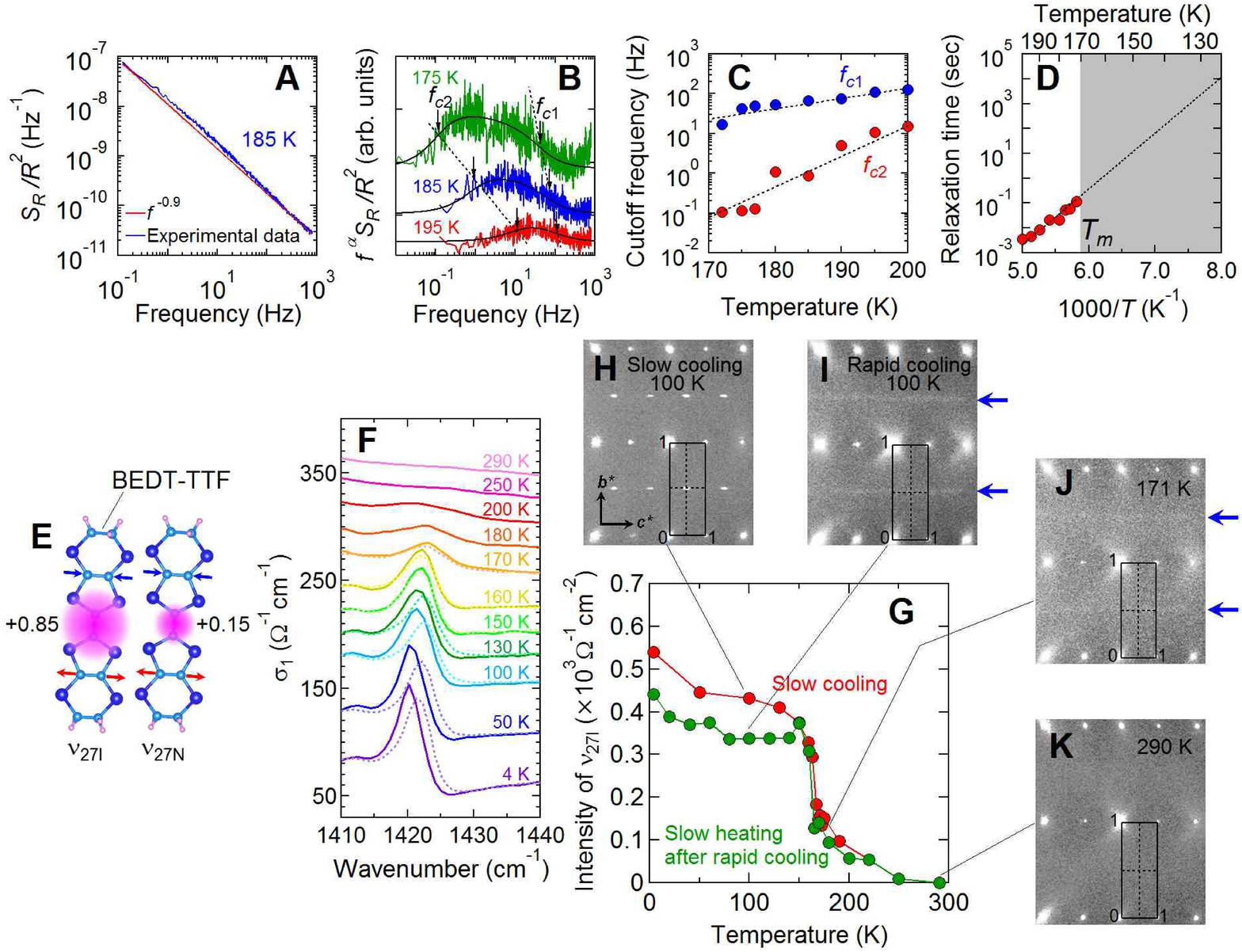}
\caption{{\bf Noise spectroscopy, optical conductivity, and x-ray diffuse scattering measurements in the charge-crystal, charge-liquid, and charge-glass states.} ({\bf A}) Typical normalized resistance noise power spectral density at 185 K. The red line indicates a fit to a background $S_R/R^2 \propto 1/f^{\alpha}$ with $\alpha=0.9$. ({\bf B}) Power spectral density multiplied by $f^{\alpha}$ at various temperatures above $T_m$. The black solid curves represent fits to the distributed Lorentzian model ({\it{11, 12}}) with a characteristic center frequency $f_c$ $=\sqrt{f_{c1} f_{c2}}$, where $f_{c1}$ and $f_{c2}$ are the high- and low-frequency cutoffs, respectively. The dashed lines are guides for the eye. ({\bf C}) Temperature dependence of the cutoff frequencies $f_{c1}$ and $f_{c2}$. The dashed lines are guides to the eye. ({\bf D}) Arrhenius plot of the relaxation time $\tau_c = 1/(2\pi f_c)$ derived from the power spectral density. The diagonal dashed line is a fit to the Arrhenius law, $\tau_0 \exp{(\frac{\Delta}{k_B T})}$, where $\tau_0$ = 10$^{-14}$ s and $\Delta/k_B$ = 5200 K. The gap size $\Delta$ corresponds to the energy scale of barriers in an energy landscape (Fig.\,3B). The charge crystallization process prevents measurements in the supercooled charge-liquid state (gray shaded area). ({\bf E}) Sketch of the infrared active vibrational mode $\nu_{27}$ of the BEDT-TTF molecule. The center frequency can be expressed as $\nu_{27}(\rho_c) = 1398 \rm{\,cm^{-1}}+ 140(1-\rho_c) \rm{\,cm^{-1}}$ ({\it{21}}). The components coming from charge-rich ($\nu_{\rm{27I}}$) and charge-poor ($\nu_{\rm{27N}}$) sites are observed at 1420 cm$^{-1}$ and 1515 cm$^{-1}$, respectively ({\it{25}}). ({\bf F}) Temperature dependence of the $\nu_{\rm{27I}}$ mode measured during slow cooling (solid lines) and slow heating after rapid cooling (dashed lines). ({\bf G}) Temperature dependence of the $\nu_{\rm{27I}}$ mode intensity. An upturn observed above $\sim$120 K during slow heating after rapid cooling is attributed to charge crystallization. ({\bf H} and {\bf I}) Oscillation photographs of the ${\it b^{\ast}}$-${\it c^{\ast}}$ plane measured at 100 K after slow cooling (H) and rapid cooling (I). In the charge-crystal state (H), clear satellite peaks appear at $\bm q_0 = (1/2\,1/2)$. By contrast, in the charge-glass state (I), diffuse lines of $\bm q_d = (1/2\, l)$ are observed. 
({\bf J} and {\bf K}) Oscillation photographs of the ${\it b^{\ast}}$-${\it c^{\ast}}$ plane measured at 171 K and 290 K, respectively. At room temperature, only Bragg reflections exist. In contrast, diffuse lines at $\bm q_d = (1/2\, l)$ are observed above $T_m$. The blue arrows in (I) and (J) indicate the diffuse lines.
} 
\end{figure*}
%%%%%%%%%%%%%%%%%%%%%%FIG 2%%%%%%%%%%%%%%%

%4
To clarify the origin of charge-glass formation in $\theta_m$-TlZn, we performed resistance noise measurements, which are a powerful probe to detect the slow dynamics associated with electronic glassiness ({\it{11, 12, 18}}). Figure\,2A shows a typical normalized noise power spectral density of the resistance fluctuations $S_R/R^2$. The baseline of $S_R/R^2$ fits well to generic $1/f$ with a slightly deviating frequency exponent, yielding  $1/f^{\alpha}$ with $\alpha \sim $ 0.8 to 0.9. For clarity, $f^{\alpha} \times S_R/R^2$ is plotted in Fig.\,2B, which clearly shows that the resistance fluctuations exhibit a broad peak structure. The data are well fitted to the distributed Lorentzian model ({\it{11, 12}}) with a characteristic center frequency $f_c$ ($=\sqrt{f_{c1} f_{c2}}$, where $f_{c1}$ and $f_{c2}$ are the high- and low-cutoff frequencies, respectively), from which we derived the temperature evolution of the relaxation time $\tau_c = 1/(2\pi f_c)$. The peak structure in $f^{\alpha} \times S_R/R^2$ becomes broader and more asymmetric with decreasing temperature (Fig.\,2, B and C), showing that the dynamics become more heterogeneous. In addition, $\tau_c$ slows drastically over several orders of magnitude (Fig.\,2D); extrapolating our data under the assumption that $\tau_c$ obeys an Arrhenius law [as observed for $\theta$-CsZn ({\it{12}}) and as expected from recent Monte Carlo simulations ({\it{15}})], we find that $\tau_c$ may be as high as $10^2$ s around $T_g$. These results indicate the emergence of slow dynamics accompanied by increasing dynamic heterogeneity upon approaching the charge glass transition. We note that the charge vitrification in the present case is distinctly different from the drastic slowing down of charge carrier dynamics and onset of non-Gaussian fluctuations observed in noise measurements as a precursor of metal-insulator transitions (MITs) ({\it{18}}). Electronic glassiness in MIT systems seemingly only becomes stabilized by disorder in the presence of strong electronic correlations
({\it{18, 19}}) and is not observed for clean samples.

%5\UTF{009B}\UTF{008D}\UTF{0093}\UTF{008A}\UTF{008B}\UTF{0081}\UTF{0081}\UTF{0095}\UTF{0082}\UTF{008C}\UTF{0082}\UTF{008B}\UTF{008B}\UTF{0095}\UTF{009A}\UTF{0084}\UTF{008D}\UTF{009D}\UTF{0087}\UTF{0080}\UTF{0081}\UTF{0081}\UTF{0081}\UTF{0084}\UTF{0081}\UTF{009A}%
Clarifying the relationship between the heterogeneous slow dynamics and local charge configurations may be key to the understanding of charge-glass formation ({\it{3, 20}}). To this end, we investigated the imbalance of charge distribution---that is, charge disproportionation---at a microscopic level by means of a charge-sensitive vibrational mode $\nu_{27}$ of the BEDT-TTF molecule. The $\nu_{27}$ mode is known as a local probe of the molecular charge $\rho_c$ ({\it{21}}) and splits into two modes, $\nu_{\rm{27I}}$ and $\nu_{\rm{27N}}$, in the presence of charge disproportionation between the A and B sites in the unit cell (Figs.\,1C and 2E), where the subscripts $\rm{I}$ and $\rm{N}$ denote the hole-rich and hole-poor sites, respectively. Figure\,2F shows the temperature dependence of the polarized optical conductivity spectra. A clear peak around 1420 cm$^{-1}$ is assigned to the $\nu_{\rm{27I}}$ mode ($\rho_c \sim 0.85$) ({\it{16}}). 
There is a slight difference in the peak frequency of 1.5 cm$^{-1}$ between the slow and rapid cooling processes, which corresponds to 0.01 in charge distribution on the BEDT-TTF molecule.

A sizable intensity of $\nu_{\rm{27I}}$ was observed above $T_m$ (Fig.\,2G), indicating the presence of charge disproportionation above $T_m$. Because the A and B sites are crystallographically equivalent above $T_m$ owing to the screw axis along the $b$ axis ({\it{16}}), the time-averaged charge distribution above $T_m$ is +0.5 per one BEDT-TTF molecule (Fig.\,1C). Therefore, the splitting of $\nu_{27}$ above $T_m$ implies that charge disproportionation above $T_m$ is not static but dynamically fluctuates on a time scale slower than that of the molecular vibrational $\nu_{27}$ motion. Because the intensity of $\nu_{\rm{27I}}$ is considered to reflect the volume of dynamically fluctuating charge clusters switching between the locally ordered and charge-liquid states, its increase with decreasing temperature suggests that the heterogeneous slow dynamics observed in the noise measurements are caused by the dynamically fluctuating charge clusters.

%%%%%%%%%%%%%%%%%%%%%%FIG 3%%%%%%%%%%%%%%%
\begin{figure}[t]
\includegraphics[width=0.95\linewidth]{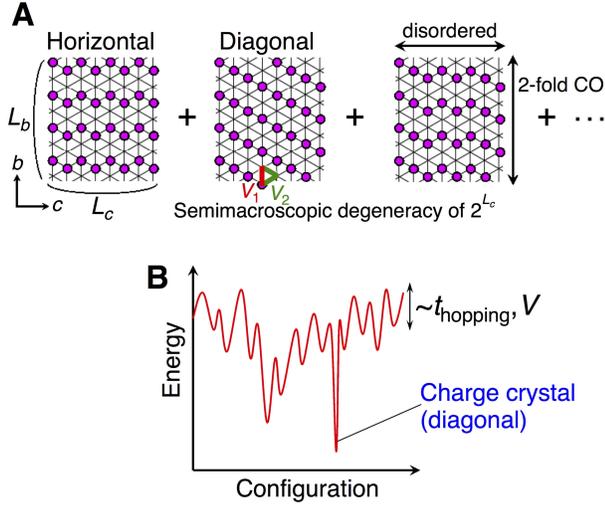}
\caption{{\bf Semimacroscopic degeneracy of striped CO patterns and energy landscape.} ({\bf A}) Schematics of various striped CO patterns. The magenta circles represent the charge-rich sites. $V_1$ and $V_2$ ($V_1>V_2$) are the nearest-neighbor Coulomb interactions. Because all these states are degenerate in the classical limit of the $t$-$V$ model, the classical ground state can be described by the superposition of these states, which has a degeneracy of $2^{L_c}$, where $L_c$ is the system length in the $c$ direction. ({\bf B}) Illustration of an energy landscape with multiple local minima separated by barriers having an energy scale of the hopping integral $t_{\rm{hopping}}$ and/or the long-range Coulomb interaction $V$, which are on the order of $\sim$ 1 eV.
}
\end{figure}
%%%%%%%%%%%%%%%%%%%%%%FIG 3%%%%%%%%%%%%%%%

%%%%%%%%%%%%%%%%%%%%%%FIG 4%%%%%%%%%%%%%%%
\begin{figure*}[t]
\includegraphics[width=0.9\linewidth]{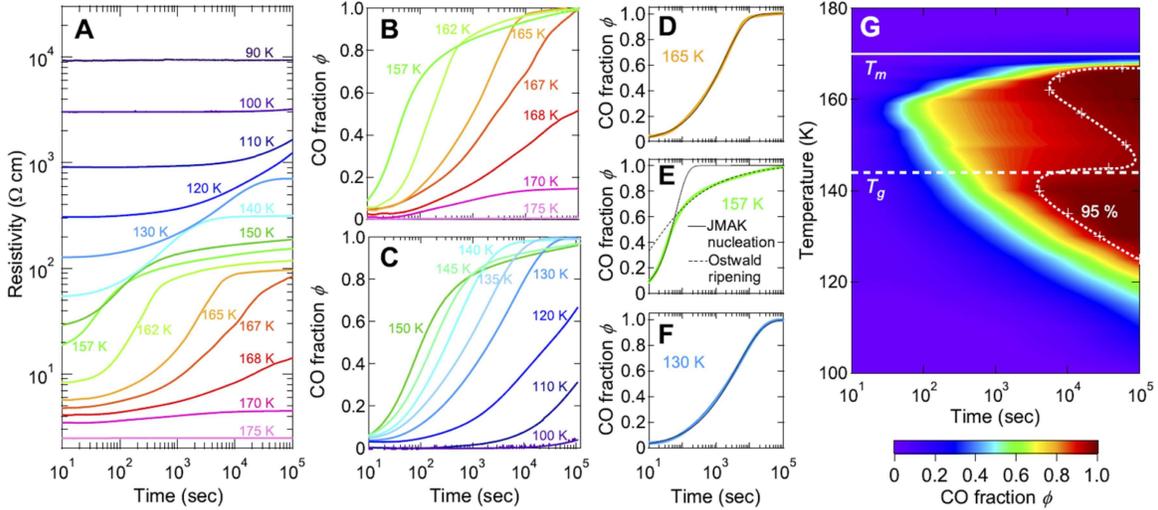}
\caption{{\bf Derivation of the time-temperature-transformation (TTT) diagram of $\theta_m$-(BEDT-TTF)$_2$TlZn(SCN)$_4$. }({\bf A}) Time-dependent resistivity change during the charge crystallization process, measured at various temperatures. ({\bf B} and {\bf C}) Time evolution of the CO volume fraction $\phi(t)$ calculated from the data in (A) using the effective medium percolation theory. Shown are the evolutions above (B) and below (C) the nose temperature.
({\bf D} to {\bf F}) $\phi(t)$ at (d) 165 K, (e) 157 K, and (f) 130 K, respectively. Also shown in (D) and (F) are fits by the JMAK formula, $\phi(t) = 1-\exp{(-kt^n)}$, where $k$ and $n$ are the JMAK parameters, and the Ostwald ripening process, $\phi(t) = 1-(1+k^{'}t)^{-1/3}$, where $k^{'}$ is a constant. ({\bf G}) TTT diagram derived from $\phi(t)$ in (B) and (C). The dotted curve connects the data points where $\phi(t)=0.95$.
} 
\end{figure*}
%%%%%%%%%%%%%%%%%%%%%%FIG 4%%%%%%%%%%%%%%% 

%6
To obtain insights into the structural origin of the heterogeneous slow dynamics, we performed x-ray diffuse scattering measurements. Oscillation photographs measured at various temperatures are shown in Fig.\,2, H to K. At room temperature, only Bragg reflections are observed (Fig.\,2K), whereas in the charge-crystal state, clear satellite peaks appear at $\bm q_0 = (1/2\,1/2)$, compatible with the diagonal CO pattern (Fig.\,2H), which is attributed to the periodicity of charge-rich and charge-poor sites accompanied by a periodical change of the C=C double bond length of the BEDT-TTF molecules. Theoretical calculations for the $\theta$-type materials based on the EHM ({\it{22}}) have suggested that the diagonal CO pattern is most stable when $V_1 > V_2$ (where $V_1$ and $V_2$ are the nearest-neighbor Coulomb interactions), which is consistent with the observations in the charge-crystal state of $\theta_m$-TlZn. In contrast, in the charge-glass state, diffuse lines at $\bm q_d = (1/2\, l)$ are observed (Fig.\,2I). The diffuse lines can be ascribed to geometric frustration.

The classical ground states of the $t$-$V$ model, which is the spinless version of the EHM, on an isosceles triangular lattice are known to be disordered owing to geometric frustration when $V_1 \geq V_2$ ({\it{23--25}}). For $V_1 = V_2$, the ground state is ``macroscopic" disordered with a degeneracy of $\sim 2^{N-1}$ (where $N$ is the number of lattice sites). On the other hand, for $V_1>V_2$, $V_1$ preferentially determines the two-fold periodic striped CO pattern along the $b$ direction, but the geometric frustration along the diagonal directions arising from the isosceles triangular lattice gives rise to a ``semimacroscopic" degeneracy of $2^{L_c}$, where $L_c$ is the system length in the $c$ direction (Fig.\,3A). However, introducing a small quantum hopping term or a long-range Coulomb potential lifts the degeneracy, which drives the system to the diagonal CO pattern ({\it{15, 22}}), although the degeneracy is presented in a wide temperature range above the ordering temperature ({\it{15}}). This situation may induce an energy landscape with multiple local minima, as illustrated in Fig.\,3B---that is, a metastable state with an amorphous stripe-glass structure as proposed in ({\it{15}})---which in turn causes the heterogeneous slow dynamics in $\theta_m$-TlZn.
Indeed, frustration is a key concept for understanding glass transitions in a variety of systems ({\it{3}}). For example, crystallization in metallic glasses is prevented if locally favored structures such as icosahedral order do not match the symmetry of the system ({\it{3, 26}}). Likewise, in $\theta_m$-TlZn, locally favored short-range electronic ordering with $\bm q_d = (1/2\, l)$ induced by geometric frustration may hinder long-range CO with $\bm q_0 = (1/2\,1/2)$, thereby causing the slow dynamics. Our results provide an experimental demonstration of recent theoretical considerations that frustration, in combination with strong quantum effects, plays an important role in the realization of quantum charge-glass states in clean systems, essentially free from impurities and defects ({\it{27--30}}).

%8.
We next examine the charge crystallization process in detail to clarify the relationship between crystallization and vitrification of electrons in $\theta_m$-TlZn. Figure\,4A displays the time evolution of the resistivity during the charge crystallization process from the supercooled charge-liquid or charge-glass state ({\it{25}}). The magnitude of the resistivity, which is a measure of the crystallization progress, increases with time and then saturates. The relaxation time becomes faster with decreasing temperature,
and then slower below 157 K, which is referred to as the ``nose temperature"; this characteristic
temperature dependence of the relaxation time can be explained by the theory of nucleation and
growth at a first-order liquid-crystal phase transition ({\it{25}}).

To quantitatively evaluate the CO volume fraction from the resistivity, we used the effective medium percolation theory ({\it{31}}). This theory describes a percolating current passing through an inhomogeneous mixture of conducting and insulating media ({\it{25}}). Through a generalized effective medium equation ({\it{32}}), we derived the time evolution of the CO volume fraction, $\phi(t)$, from the time-dependent resistivity data at various temperatures (Fig.\,4, B and C). In the high-temperature region, $\phi(t)$ can be fitted over the whole time range by the Johnson-Mehl-Avrami-Kolmogorov (JMAK) formula describing a conventional nucleation and growth process, where $\phi(t) = 1-\exp{(-kt^n)}$ (here, $k$ and $n$ are the JMAK parameters) ({\it{25, 33}}) (Fig.\,4E). By contrast, near the nose temperature, $\phi(t)$ can be fitted to the JMAK formula only in the early stage of crystal growth (Fig.\,4E). For later times, a crossover to the Ostwald ripening that describes a rearrangement of crystal grain boundaries, where $\phi(t) = 1-(1+k^{'}t)^{-1/3}$ (here, $k^{'}$ is a constant) ({\it{34}}), explains the observed more moderate time evolution of $\phi(t)$. Such a process is often observed in the final stage of crystal growth ({\it{25}}). Interestingly, below 145 K, $\phi(t)$ again exhibits a steep increase over the whole time range, which can be fitted to the JMAK formula (Fig.\,4F); the temperature of 145 K is close to $T_g$. Crystallization below $T_g$ is studied in many fields of materials science, and an enhancement of the crystallization rate at $T_g$ has been reported ({\it{35, 36}}). The origin has been discussed, for example, in terms of a crystal--glass interface ({\it{35, 36}}). In this scenario, the volume contraction upon crystallization below $T_g$ provides free volume for atoms or molecules surrounding the crystal, which leads to a mobility increase at the crystal--glass interface, resulting in an enhancement of the crystal growth rate at $T_g$.
We speculate that the same surface dynamics occurs at the charge crystal--glass interface in $\theta_m$-TlZn. 

%9
Figure\,4G displays the contour map of the CO fraction plotted in the time-temperature plane [a so-called time-temperature-transformation (TTT) diagram]. The obtained TTT diagram clearly reflects the two characteristic features discussed above: the nose structure around 160 K, and the enhancement of crystal growth close to $T_g$. These observations suggest that the crystallization process of electrons in solids can be described by the nucleation and growth process of a liquid---as observed in conventional glass-forming liquids such as structural and metallic glasses ({\it{35--38}})---and that charge crystallization and vitrification are closely related.

Our study reveals that when electrons in a strongly correlated system are put on a geometrically frustrated lattice, their dynamics are similar to those known from structural relaxation in conventional glass-forming liquids, although the additional role of the lattice degrees of freedom for charge-glass formation on a geometrically frustrated system requires further investigation. The convenient time and temperature scales of the present material and the possibility of inferring volume fractions from easily accessible charge transport will enable investigations of aging, memory effects, cooperativity, and the presence or absence of an underlying true phase transition from a different perspective.

\section*{Acknowledgments}
We thank K. Yoshimi, T. Kato, M. Naka, H. Shiba, C. Hotta, and K. Yamamoto for fruitful discussions, and J. Kudo and M. Kurosu for technical assistance. Synchrotron radiation measurements were performed at SPring-8 with the approvals of the Japan Synchrotron Radiation Research Institute (2014B1340, 2014B1752, 2015A1777, 2015B1752, 2015B1756, 2016A0073, and 2016B0073). X-ray diffraction study was performed under the approval of the Photon Factory Program Advisory Committee (proposal 2014S2-001). Supported by the Deutsche Forschungsgemeinschaft within the Transregional Collaborative Research Center SFB/TR49 (J.M.); Grants-in-Aid for Scientific Research (grants 24340074, 25287080, 26610096, 26102014, 15H00984, 15H00988, 15K13511, 15K17688, 16H04010, 16K05430, 16K05744, 17H05138, and 17H05143) from MEXT and JSPS; a Grant-in-Aid for Scientific Research on Innovative Areas ``p-Figuration" (grant 26102001); and the Canon Foundation.

%%%%%%%%%References


\begin{thebibliography}{99}

\bibitem{Anderson95}
P.\,W. Anderson, {\it Science} {\bf 267}, 1615 (1995).

%glass review
\bibitem{Debenedetti01}
P.\,G. Debenedetti, F.\,H. Stillinger, {\it Nature} {\bf 410}, 259--267 (2001).

\bibitem{Shintani06}
H. Shintani, H. Tanaka, {\it Nature Phys.} {\bf 2}, 200--206 (2006).

\bibitem{Kawasaki10}
T. Kawasaki, H. Tanaka, {\it Proc. Natl. Acad. Sci. USA} {\bf 107}, 14036--14041 (2010).

\bibitem{Angell88}
C.\,A. Angell, {\it J. Phys. Chem. Solids} {\bf 49}, 863-871 (1988).

\bibitem{Mori98-1}
H. Mori, S. Tanaka, T. Mori, A. Kobayashi, H. Kobayashi, {\it Bull. Chem. Soc. Jpn.}  {\bf 71}, 797--806 (1998).

\bibitem{Mori98-2}
H. Mori, S. Tanaka, T. Mori, {\it Phys. Rev. B} {\bf 57}, 12023--12029 (1998).

\bibitem{Sawano05}
F. Sawano {\it et al.}, {\it Nature} {\bf 437}, 522 (2005).

%CO theory
\bibitem{Seo00}
H. Seo, {\it J. Phys. Soc. Jpn.} {\bf 69}, 805--820 (2000).

\bibitem{Watanabe04}
M. Watanabe, Y. Noda, Y. Nogami, H. Mori, {\it J. Phys. Soc. Jpn.} {\bf 73}, 116--122 (2004).

%Kagawa group
\bibitem{Kagawa13}
F. Kagawa {\it et al.}, {\it Nature Phys.} {\bf 9}, 419--422 (2013).

\bibitem{Sato14-1}
T. Sato {\it et al.}, {\it Phys. Rev. B} {\bf 89}, 121102(R) (2014).

%Kagawa group
\bibitem{Sato14-2}
T. Sato {\it et al.}, {\it J. Phys. Soc. Jpn} {\bf 83}, 083602 (2014).

\bibitem{Oike15}
H. Oike {\it et al.}, {\it Phys. Rev. B} {\bf 91}, 041101(R) (2015). 

\bibitem{Mahmoudian15}
S. Mahmoudian {\it et al.}, {\it Phys. Rev. Lett.} {\bf 115}, 025701 (2015).

\bibitem{Suzuki04}
K. Suzuki, K. Yamamoto, K. Yakushi, {\it Phys. Rev. B} {\bf 69}, 085114 (2004).

\bibitem{Udagawa07}
M. Udagawa, Y. Motome, {\it Phys. Rev. Lett.} {\bf 98}, 206405 (2007).

\bibitem{Jaroszynski02}
J. Jaroszy$\acute{\rm{n}}$ski, D. Popovi$\acute{\rm{c}}$, T. M. Klapwijk, {\it Phys. Rev. Lett.} {\bf 89}, 276401 (2002).

\bibitem{Dobrosavljevic03}
V. Dobrosavljevi$\acute{\rm{c}}$, D. Tanaskovi$\acute{\rm{c}}$, A. A. Pastor, {\it Phys. Rev. Lett.} {\bf 90}, 016402 (2003).

\bibitem{Ediger12}
M.\,D. Ediger, P. Harrowell,  {\it J. Chem. Phys.} {\bf 137}, 080901 (2012).

\bibitem{Yamamoto05}
T. Yamamoto {\it et al.}, {\it J. Phys. Chem B} {\bf 109}, 15226--15235 (2005).

\bibitem{Naka14}
M. Naka, H. Seo, {\it J. Phys. Soc. Jpn} {\bf 83}, 053706 (2014).

\bibitem{Hotta06}
C. Hotta, N. Furukawa, {\it Phys. Rev. B} {\bf 74}, 193107 (2006).

\bibitem{Han08}
Y. Han {\it et al.}, {\it Nature} {\bf 456}, 898--903 (2008).

\bibitem{Supplementary}
See supplementary materials.

\bibitem{Hirata13}
A. Hirata {\it et al}., {\it Science} {\bf 341}, 376--379 (2013).

\bibitem{Dagotto05}
E. Dagotto, {\it Science} {\bf 309}, 257--262 (2005).

\bibitem{Schmalian00}
J. Schmalian, P.\,G. Wolynes, {\it Phys. Rev. Lett.} {\bf 85}, 836--839 (2000).

\bibitem{Jamei05}
R. Jamei, S. Kivelson, B. Spivak, {\it Phys. Rev. Lett.} {\bf 94}, 056805 (2005).

%Theory(glassy state)
\bibitem{Yoshimi12}
K. Yoshimi, H. Maebashi, {\it J. Phys. Soc. Jpn} {\bf 81}, 063003 (2012).

\bibitem{Stauffer94}
D. Stauffer, A. Aharony, 
{\it Introduction to Percolation Theory.} Taylor and Francis, London, second edition, (1994).

\bibitem{McLachlan90}
D.\,S. McLachlan, M. Blaszkiewicz, R.\,E. Newnham, {\it J. Am. Ceram. Soc.} {\bf 73}, 2187--2203 (1990).

\bibitem{Avrami39}
M. Avrami, {\it J. Chem. Phys.} {\bf 7}, 1103 (1939).

\bibitem{Lifshitz61}
I.\,M. Lifshitz, V.\,V. Slyozov, {\it J. Phys. Chem. Solids.} {\bf 19}, 35--50 (1961).


\bibitem{Konishi07}
T. Konishi, H. Tanaka, {\it Phys. Rev. B} {\bf 76}, 220201(R) (2007).

\bibitem{Sun11}
Y. Sun, L. Zhu, K.\,L. Kearns, M.\,D. Ediger, L. Yu, {\it Proc. Natl. Acad. Sci. USA} {\bf 108}, 5990--5995 (2011).

\bibitem{Kim96}
Y.\,J. Kim, R. Busch, W.\,L. Johnson, A.\,J. Rulison, W.\,K. Rhim, {\it Appl. Phys. Lett.} {\bf 68}, 1057--1059 (1996).

\bibitem{Moore11}
E.\,B. Moore, V. Molinero, {\it Nature} {\bf 479}, 506--508 (2001).


\end{thebibliography}
\end{document}